\documentclass[usenatbib]{mnras}

\usepackage{graphicx}	
\usepackage[fleqn]{amsmath}	
\usepackage{amssymb}	
\usepackage{multicol}        
\usepackage{bm}		
\usepackage[T1]{fontenc}
\usepackage{ae,aecompl}

\usepackage{newtxtext,newtxmath}

\usepackage{xcolor,xspace}
\usepackage{booktabs}  
\usepackage{morefloats}  

\usepackage[switch,mathlines]{lineno}  

\usepackage{enumerate} 

\usepackage{soul}

\definecolor{linkcolor}{rgb}{0, 0, 1.}			
\definecolor{funcolor}{rgb}{0.65, 0.16, 0.16}	
\definecolor{parcolor}{HTML}{1E8449}        
\definecolor{bg_code}{rgb}{0.95,0.95,0.95}		

\newcommand{\tbref}[1]{Table~\ref{#1}}
\newcommand{\figref}[1]{Figure~\ref{#1}}

\newcommand{\alphamm}{\alpha_\mathrm{0.89-3.1\,mm}}
\newcommand{\MSun}{M_\odot}

\newcommand{\mJybeam}{\,mJy~beam$^{-1}$\xspace}
\newcommand{\muJybeam}{\,$\mu$Jy~beam$^{-1}$\xspace}

\newcommand{\Rcore}{R_\mathrm{core}}

\newcommand{\amax}{a_\mathrm{max}}
\newcommand{\microJybeam}{$\mu$Jy\,beam$^{-1}$}

\newcommand{\revdel}[1]{}  

\hypersetup{
draft=true,
colorlinks=true,
linkcolor=linkcolor,
citecolor=linkcolor,
filecolor=linkcolor,
urlcolor=linkcolor}

\hypersetup{
pdfauthor={M. Tazzari},
pdftitle={The first ALMA survey of protoplanetary discs at 3 mm: demographics of grain growth in the Lupus region},
pdfkeywords={pdf keywords},
bookmarksnumbered=true,
draft=false}

\title[The first ALMA protoplanetary disc survey at 3\,mm]{
The first ALMA survey of protoplanetary discs at 3\,mm: demographics of grain growth in the Lupus region
}

\author[M. Tazzari et al.]{M. Tazzari$^{1}$\thanks{Contact e-mail: \href{mailto:mtazzari@ast.cam.ac.uk}{mtazzari@ast.cam.ac.uk}},
L. Testi$^{2,3}$,
A. Natta$^{4}$, 
J. P. Williams$^{5}$,
M. Ansdell$^{6,7}$,
J. M. Carpenter$^{8}$,\newauthor
S. Facchini$^{2}$,
G. Guidi$^{9}$,
M. Hogherheijde$^{10}$,
C. F. Manara$^{2}$,
A. Miotello$^{2}$,
N. van der Marel$^{11}$
\\
$^{1}$Institute of Astronomy, University of Cambridge, Madingley Road, CB3 0HA,  Cambridge, UK\\
$^{2}$European Southern Observatory, Karl-Schwarzschild-Str. 2, D-85748 Garching, Germany\\
$^{3}$INAF/Osservatorio Astrofisico di Arcetri, Largo E. Fermi 5, I-50125 Firenze, Italy\\
$^{4}$School of Cosmic Physics, Dublin Institute for Advanced Studies, 31 Fitzwilliam Place, Dublin 2, Ireland\\
$^{5}$Institute for Astronomy, University of Hawai‘i at Manoa,2680 Woodlawn Dr., Honolulu, HI, USA\\
$^{6}$Flatiron Institute, Simons Foundation, 162 Fifth Ave, New York, NY 10010, USA\\
$^{7}$NASA Headquarters, 300 E Street SW, Washington, DC 20546, USA\\
$^{8}$Joint ALMA Observatory, Avenida Alonso de C\'ordova 3107, Vitacura, Santiago, Chile\\
$^{9}$ETH Zurich, Institute for Particle Physics and Astrophysics, Wolfgang-Pauli-Str. 27, 8093 Zurich, Switzerland\\
$^{10}$Leiden Observatory, Leiden University, P.O. Box 9531, NL-2300 RA Leiden, the Netherlands\\
$^{11}$Physics \& Astronomy Department, University of Victoria, 3800 Finnerty Road, Victoria, BC, V8P 5C2, Canada
}

\date{\today}

\pubyear{2020}

\begin{document}
\label{firstpage}
\pagerange{\pageref{firstpage}--\pageref{lastpage}}
\maketitle

\begin{abstract}
We present the first ALMA survey of protoplanetary discs at 3\,mm, targeting 36 young stellar objects in the Lupus star-forming region with deep observations (sensitivity 20-50\,$\mu$Jy\,beam$^{-1}$) at $\sim0.35\arcsec$ resolution ($\sim$50\,au).
Building on previous ALMA surveys at 0.89 and 1.3\,mm that observed the complete sample of Class~II discs in Lupus at a comparable resolution, we aim to assess the level of grain growth in the relatively young Lupus region. 
We measure 3\,mm integrated fluxes, from which we derive disc-averaged 1-3\,mm spectral indices. 
We find that the mean spectral index of the observed Lupus discs is $\alpha_\mathrm{1-3\,mm}=2.23\pm0.06$, in all cases $\alpha_\mathrm{1-3\,mm}<3.0$, with a tendency for larger spectral indices in the brightest discs and in transition discs. 
Furthermore, we find that the distribution of spectral indices in Lupus discs is statistically indistinguishable from that of the Taurus and Ophiuchus star-forming regions. 
Assuming the emission is optically thin, the low values $\alpha_\mathrm{1-3\,mm}\leq 2.5$ measured for most discs can be interpreted with the presence of grains larger than 1\,mm.
The observations of the faint discs in the sample can be explained without invoking the presence of large grains, namely through a mixture of optically thin and optically thick emission from small grains. However, the bright (and typically large) discs do inescapably require the presence of millimeter-sized grains in order to have realistic masses. 
Based on a disc mass argument, our results challenge previous claims that the presence of optically thick sub-structures may be a universal explanation for the empirical millimeter size-luminosity correlation observed at 0.89\,mm.
\end{abstract}



\section{Introduction}
Protoplanetary discs are the birth place of planets. In the core-accretion scenario \citep{safronov:1972,Wetherill:1980lr}, the first step to form a terrestrial planet is the growth of the typically micron-sized interstellar medium (ISM) dust grains to millimeter and centimeter-sized pebbles. Although this initial growth is favoured in the dense protoplanetary disc mid-planes, evidence of growth at early stages has been obtained also in very young (Class~0~and~I) systems \citep{Miotello:2014aa,Agurto-Gangas:2019aa,Galametz:2019aa}. Planet formation then proceeds with the assembly of kilometer-sized planetesimals, which eventually form rocky planets and the cores of giant planets. 
Thanks to their sensitivity to the thermal emission of dust grains, observations at sub-millimeter and millimeter wavelengths are a key probe of the first stages of the planet formation process \citep[][and references therein]{Testi:2014kx}. 

In a smooth protoplanetary disc, millimeter and centimeter sized grains are expected to undergo fast inward radial drift due to the aerodynamic drag exerted on them by the gas in sub-Keplerian motion \citep{Weidenschilling:1977lr}. Since radial drift occurs very quickly compared to the disc dynamical timescale, discs should be thoroughly depleted of large grains within the first 1\,Myr of their life \citep{Brauer:2008kq}. 

Observations, however, are in contrast with this scenario. As known from seminal 
sub-millimeter and millimeter photometric observations \citep[e.g.,][]{Beckwith:1990qf}, and 
recently confirmed by ALMA spatially resolved observations \citep[e.g.,][]{2041-8205-808-1-L3, Ansdell:2016qf,Isella:2016ww,Andrews:2018aa}, a large fraction of protoplanetary discs are bright 
and relatively extended (between 30 and 100\,au in radius; see \citealt{Hendler:2020aa})
when observed at 1\,mm (a wavelength most sensitive to the thermal emission of millimeter sized grains). 
The measured disc fluxes and extents at 1\,mm suggest that radial drift has to be effectively slowed down or even halted in most discs \citep{Birnstiel:2010aa,Pinilla:2012xy}.
Thanks to their ability to trap (or, at least, decelerate the inward drift of) dust grains, local maxima in the gaseous component of a protoplanetary disc are a ready solution to the radial drift problem.
This scenario is supported by the recent high angular resolution ALMA observations  \citep[e.g.,][]{Isella:2016ww,Fedele:2018aa,Clarke:2018aa,Andrews:2018aa,Long:2018aa,Long:2019aa} which showed a suggestive recurrence in the discs millimetre continuum emission of orderly and axisymmetric structures at small spatial scales. Whether the rings observed in the continuum emission are effective dust traps is a topic of active research \citep{Dullemond:2018aa}. The origin of local maxima can be linked to a variety of physical mechanisms, such as the interaction with an embedded forming planet \citep[e.g., ][]{Pinilla:2012aa,Clarke:2018aa}, zonal flows \citep{Johansen:2005aa,Flock:2015aa}, and the presence of vortices \citep{1995A&A...295L...1B,Klahr:1997fk}.

The optical properties of dust grains can be used to probe their spatial distribution in discs: as grains grow to sizes close to the observing wavelength, their opacity changes significantly, leaving a signature of their presence in the disc spectral energy distribution (SED). Specifically, at millimeter and centimeter wavelengths, the spectral index of the emission of optically thin dust can be directly linked, for a given temperature, to the spectral dependence of the dust opacity coefficient, which in turn depends on the maximum grain size of the emitting dust \citep{Natta:2007ye,Draine:2006ty}. 

Extensive observational studies investigating the level of grain growth in discs have been conducted in the Taurus-Auriga \citep{Andrews:2005fk, Rodmann:2006hl,Ricci:2010fv} and Ophiuchus \citep{Andrews:2007aa,Ricci:2012aa} star-forming regions. Other studies, with more limited sample sizes have also targeted southern star-forming regions  such as Lupus and Chamaeleon \citep{Lommen:2007aa,Ubach:2012ul} and the distant Orion Nebula Cluster \citep{Mann:2010aa, Ricci:2011aa}. In the majority of discs these studies found relatively small millimeter spectral indices ($\alphamm\sim2.5$), which were interpreted (by means of simplified disc modelling) in terms of emission from millimeter-sized grains. 
It is noteworthy that all these surveys aimed at measuring the spectral index of the dust opacity absorption coefficient are by no means complete in any star-forming region, and in all cases are spatially unresolved. 

Evidence of enhanced grain growth in the inner disc region in line with the expectations for radial drift was obtained by \citet{Perez:2012fk,Perez:2015fk} and \citet{Tazzari:2016qy}, and more recently by \citet{Tripathi:2018aa}, \citet{Huang:2018aa,Huang:2020aa}, \citet{Carrasco-Gonzalez:2019aa}, and \citet{Long:2020ab} for a dozen of discs 
in the Taurus-Auriga and Ophiuchus region for which spatially resolved observations in a wide wavelength range (0.88\,mm to 1\,cm) were available. 
Although the self-consistent modelling of disc structure and dust emission implemented in these studies constitutes a refinement over previous works and the evidence for grain size variations is robustly demonstrated, the statistical relevance of the results is limited by the small  sample of discs with homogeneous observations in the 0.89-3\,mm wavelength range.

In recent years, ALMA has been used to perform extensive surveys of Class~II discs in several star-forming regions such as Lupus \citep{Ansdell:2016qf,Ansdell:2018aa}, Chamaeleon~I \citep{Pascucci:2016aa}, $\sigma$~Ori \citep{Ansdell:2017aa}, Upper~Scorpius \citep{Barenfeld:2016lr}, Ophiuchus \citep{Cox:2017aa,Cieza:2019aa}, Taurus \citep{Akeson:2014aa,Akeson:2019aa,Long:2018aa,Long:2019aa}, IC~348 \citep{Ruiz-Rodriguez:2018aa}, the Orion Nebula Cluster \citep{Eisner:2018aa}, and NGC~2024 \citep{van-Terwisga:2020aa}.
Although these studies provided a new wealth of information on the discs mass, size, gas-to-dust mass ratio, and spatial brightness distribution, they only probed the disc emission at 0.89 or 1.3\,mm. Extensive observations at longer wavelength are needed to assess the level of grain growth in these regions.

Here we present the first ALMA survey of protoplanetary discs at 3\,mm, which targeted a sample of 36 objects (38\% of all Lupus Class~II discs), covering the brightest 55\% of the Class~II discs previously detected at 0.89\,mm \citet{Ansdell:2016qf}.
The moderate angular resolution ($\sim0.35\arcsec$) and high sensitivity (20-50\muJybeam) allowed us to detect all the discs at 3\,mm and to resolve the largest ones.

In this paper we focus on the spatially-integrated analysis of fluxes and spectral indices, discussing the implications on the grain properties and on empirical relations such as the millimeter continuum size-luminosity relation \citep{Tripathi:2017aa}. 
A forthcoming paper \citep{Tazzari:2020ab} will perform a homogeneous analysis of the multi-wavelength observations that are available for these Lupus discs  at 0.89, 1.3, and 3\,mm, with a particular focus on the disc sizes.

The paper is organised as follows.
In Section~\ref{sect:sample} we introduce the source sample and in Section~\ref{sect:observations} we describe observational setup and calibration details. 
In Section~\ref{sect:results} we present the measured 3\,mm fluxes and the inferred $\alphamm$ spectral indices, with a comparison with other regions. 
In Section~\ref{sect:discussion} we discuss the results, presenting the implications for the level of grain growth in Lupus, and discussing the new insights on the interpretation of the millimeter size-luminosity relation in light of this new 3\,mm data.
In Section~\ref{sect:conclusions} we draw our conclusions.

\section{Sample}
\label{sect:sample}
In this study we present 3\,mm observations of 36 disc-bearing young stellar objects (YSOs) of the Lupus star-forming region. The objects belong to the Lupus I to IV clouds and are classified as Class~II sources \citep{Merin:2008xy} or have flat infrared (IR) excess measured between the 2MASS \textit{Ks} (2.2$\mu$m) and \textit{Spitzer} MIPS-1 (24$\mu$m) bands \citep{Evans:2009ys}. 
A near-complete census of the 0.89\,mm brightness of Lupus Class~II discs has recently been carried out with ALMA by \citet{Ansdell:2016qf}. To build the initial sample from which we select the sources for this study, we complement the \citealt{Ansdell:2016qf} sample with the 0.89\,mm ALMA observations at comparable sensitivity and resolution by \citet{MacGregor:2017aa} for Sz~75/GQ~Lup, by \citet{Cleeves:2016lr} for Sz~82/IM~Lup, and by \citet{Canovas:2016aa} for Sz~91. 
From the resulting sample of 92 Class~II discs (65 detected at 0.89\,mm), we select those that can be detected with a signal-to-noise ratio larger than 10 when observed at 3\,mm with a 0.45\arcsec resolution and a nominal sensitivity between 30 and 70\muJybeam (approximately 10 and 2 minutes on source, respectively). 
To extrapolate the predicted 3\,mm brightness, we assumed a conservative spectral index of 3.0, compatible with the spectral indices found by \citet{Lommen:2007aa} and \citet{Ubach:2012ul} in their Lupus samples. The requirement of a 3\,mm detection with signal-to-noise ratio larger than 10 ensured that the absolute uncertainty on the 0.89-1.3\,mm integrated spectral index is on average smaller than 0.12, given the sensitivity of the 0.89\,mm observations.
This selection resulted in total 36 discs, which have been detected at 0.89\,mm with a peak brightness larger than 14\mJybeam at a resolution of $\sim 0.3\arcsec$: 33 discs from \citet{Ansdell:2016qf}, plus Sz~75/GQ~Lup,  Sz~82/IM~Lup, and Sz~91. The sources are listed in \tbref{tb:YSOs}.
The sample includes 9 transition discs with reported cavities larger than 20\,au \citep{van-der-Marel:2018aa}: Sz~84, RY~Lup, MY~Lup, J16070854-3914075, Sz~91, Sz~100, J16083070-3828268, Sz~111, Sz~118, Sz~123.
\begin{table*}
\caption{ALMA protoplanetary disc survey at 3\,mm: properties of the Lupus young stellar objects in the sample.}
\begin{tabular}{llccccccc}
\midrule
\toprule
Name & Other Name & R.A. & Dec. & \multicolumn{2}{c}{Identifiers} & $d$ & $M_\star$ & Notes\\
\cline{5-6} \\[-0.3cm]
 &  & (J2015.5) & (J2015.5) & 2MASS & Gaia DR2  & (pc)  & ($M_\odot$) & \\   
   \midrule
Sz 65                &                      & 15:39:27.76    & -34:46:17.55   & J15392776-3446171    &  6013399894569703040 & $155$ & $0.70$ &  \\ 
Sz 66                &                      & 15:39:28.27    & -34:46:18.42   & J15392828-3446180    &  6013399830146943104 & $157$ & $0.29$ &  \\ 
J15450634-3417378    &                      & 15:45:06.34    & -34:17:37.83   & J15450634-3417378    & ...                  & $160$ & ... &  \\ 
J15450887-3417333    &                      & 15:45:08.86    & -34:17:33.80   & J15450887-3417333    &  6014696875913435520 & $155$ & $0.14$ &  \\ 
Sz 68                & HT Lup               & 15:45:12.85    & -34:17:30.98   & J15451286-3417305    &  6014696841553696768 & $154$ & $2.15$ & $^{(1)}$ \\ 
Sz 69                &                      & 15:45:17.39    & -34:18:28.64   & J15451741-3418283    &  6014696635395266304 & $154$ & $0.20$ &  \\ 
Sz 71                & GW Lup               & 15:46:44.71    & -34:30:36.04   & J15464473-3430354    &  6014722194741392512 & $155$ & $0.41$ & $^{(2)}$ \\ 
Sz 72                &                      & 15:47:50.61    & -35:28:35.76   & J15475062-3528353    &  6011573266459331072 & $155$ & $0.37$ &  \\ 
Sz 73                &                      & 15:47:56.93    & -35:14:35.14   & J15475693-3514346    &  6011593641784262400 & $156$ & $0.78$ &  \\ 
Sz 74                &                      & 15:48:05.22    & -35:15:53.30   & J15480523-3515526    &  6011581856393989120 & $150$ & $0.30$ &  \\ 
Sz 75                & GQ Lup               & 15:49:12.09    & -35:39:05.42   & J15491210-3539051    &  6011522757643074304 & $151$ & $0.78$ &  \\ 
Sz 82                & IM Lup               & 15:56:09.19    & -37:56:06.49   & J15560921-3756057    &  6010135758090335232 & $158$ & $0.95$ & $^{(2)}$ \\ 
Sz 83                & RU Lup               & 15:56:42.30    & -37:49:15.83   & J15564230-3749154    &  6010114558131195392 & $159$ & $0.67$ & $^{(2)}$ \\ 
Sz 84                &                      & 15:58:02.50    & -37:36:03.09   & J15580252-3736026    &  6010216537834709760 & $152$ & $0.17$ &  \\ 
Sz 129               &                      & 15:59:16.46    & -41:57:10.66   & J15591647-4157102    &  5995168724780802944 & $161$ & $0.78$ & $^{(2)}$ \\ 
J15592838-4021513    & RY Lup               & 15:59:28.37    & -40:21:51.59   & J15592838-4021513    &  5996151172781298304 & $158$ & $1.53$ & $^{(1)}$ \\ 
J16000236-4222145    &                      & 16:00:02.34    & -42:22:14.96   & J16000236-4222145    &  5995139484643284864 & $163$ & $0.23$ &  \\ 
J16004452-4155310    & MY Lup               & 16:00:44.50    & -41:55:31.29   & J16004452-4155310    &  5995177933191206016 & $156$ & $1.09$ & $^{(2)}$ \\ 
J16011549-4152351    &                      & 16:01:15.49    & -41:52:35.19   & J16011549-4152351    & ...                  & $160$ & ... &  \\ 
Sz 133               &                      & 16:03:29.37    & -41:40:02.17   & J16032939-4140018    &  5995094095435598848 & $155$ & ... &  \\ 
J16070854-3914075    &                      & 16:07:08.54    & -39:14:07.89   & J16070854-3914075    &  5997076721058575360 & $177$ & ... &  \\ 
Sz 90                &                      & 16:07:10.06    & -39:11:03.65   & J16071007-3911033    &  5997077167735183872 & $160$ & $0.78$ &  \\ 
Sz 91                &                      & 16:07:11.57    & -39:03:47.85   & J16071159-3903475    &  5997091358307172224 & $158$ & $0.51$ &  \\ 
Sz 98                & HK Lup               & 16:08:22.48    & -39:04:46.81   & J16082249-3904464    &  5997082867132347136 & $156$ & $0.67$ &  \\ 
Sz 100               &                      & 16:08:25.75    & -39:06:01.59   & J16082576-3906011    &  5997082046818385408 & $137$ & $0.14$ &  \\ 
J16083070-3828268    &                      & 16:08:30.69    & -38:28:27.24   & J16083070-3828268    &  5997490206145065088 & $155$ & $1.53$ & $^{(1)}$ \\ 
J16083427-3906181    & V856 Sco             & 16:08:34.27    & -39:06:18.68   & J16083427-3906181    &  5997082081177906048 & $160$ & ... &  \\ 
Sz 108B              &                      & 16:08:42.87    & -39:06:15.03   & ...                  &  5997082218616859264 & $168$ & $0.17$ &  \\ 
Sz 110               &                      & 16:08:51.56    & -39:03:18.07   & J16085157-3903177    &  5997082390415552768 & $159$ & $0.23$ &  \\ 
J16085324-3914401    &                      & 16:08:53.23    & -39:14:40.53   & J16085324-3914401    &  5997033290348155136 & $167$ & $0.29$ &  \\ 
Sz 111               &                      & 16:08:54.67    & -39:37:43.50   & J16085468-3937431    &  5997006897751436544 & $158$ & $0.51$ &  \\ 
Sz 113               &                      & 16:08:57.79    & -39:02:23.21   & J16085780-3902227    &  5997457736191421184 & $163$ & $0.19$ &  \\ 
Sz 114               &                      & 16:09:01.84    & -39:05:12.79   & J16090185-3905124    &  5997410491550194816 & $162$ & $0.19$ & $^{(2)}$ \\ 
Sz 118               &                      & 16:09:48.64    & -39:11:17.21   & J16094864-3911169    &  5997405509388068352 & $163$ & $1.04$ &  \\ 
Sz 123               &                      & 16:10:51.57    & -38:53:14.13   & J16105158-3853137    &  5997416573223873536 & $162$ & $0.55$ &  \\ 
J16124373-3815031    &                      & 16:12:43.74    & -38:15:03.42   & J16124373-3815031    &  5997549820286701440 & $159$ & $0.45$ &  \\
\bottomrule
\end{tabular}
\begin{flushleft}
\textbf{Note.}
Name is the designation used in this paper (with notable alternative names where available).
Right Ascension (R.A.) and Declination (Dec.) are from Gaia DR2 \citep{Gaia-Collaboration:2018aa}. 
Distances ($d$) are computed using Gaia DR2 measurements by \cite{Bailer-Jones:2018aa}; their typical uncertainty is $\sim$2\,pc, with larger values of $13\,$pc for Sz 133 and J16070854-3914075.
J16011549-415235 and J15450634-3417378 were not found in Gaia DR2. For these sources, we assume the average distance of the Lupus cloud complex (160 pc, \citealt{Manara:2018aa}). 
Stellar masses ($M_\star$) are taken from Table A.1 in \citealt{Alcala:2019aa}: they are inferred from the UV spectroscopic measurements by \citet{Alcala:2017aa}, account for updated distances drawn from Gaia DR2, and are based on \citet{Baraffe:2015aa} pre-main-sequence evolutionary tracks, if available.
$^{(1)}$~Based on \citet{Siess:2000qy} evolutionary tracks.
$^{(2)}$~Observed in the DSHARP survey \citep{Andrews:2018aa}.
A machine-readable version of this table is available online (see the Data Availability statement).
\end{flushleft}
\label{tb:YSOs}
\end{table*}

\begin{figure}
    \centering
    \includegraphics[width=\hsize]{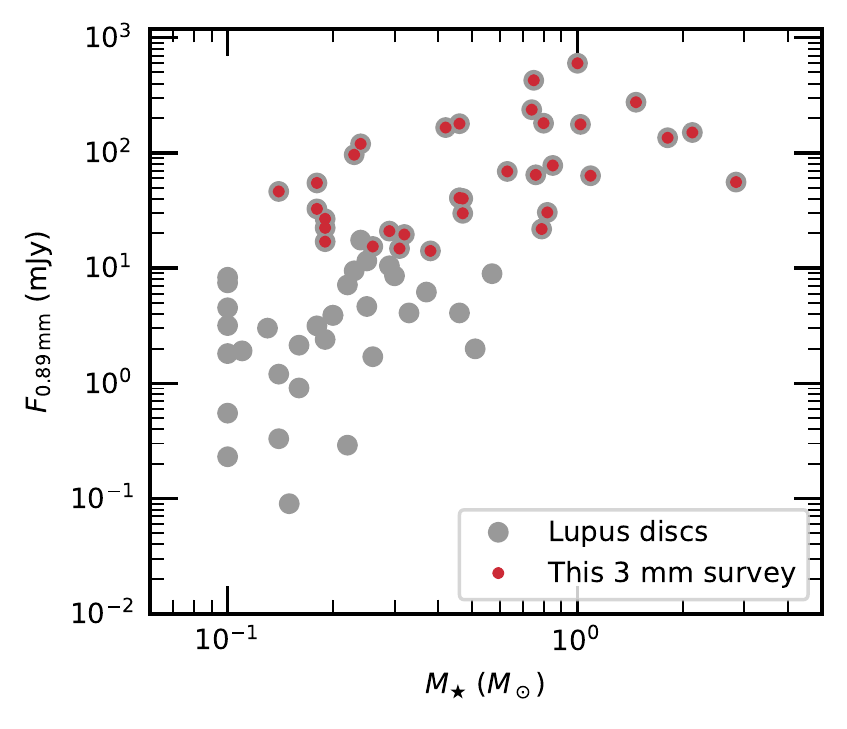}
    \caption{Overview of the sample properties: integrated 0.89\,mm flux as a function of stellar mass. The Lupus discs detected by the 0.89\,mm survey \citep{Ansdell:2016qf} plus Sz~75/GQ~Lup \citep{MacGregor:2017aa}, Sz~82/IM~Lup \citep{Cleeves:2016lr}, and Sz~91 \citep{Canovas:2016aa} that were not included in \citet{Ansdell:2016qf} are shown in grey circles. Discs targeted by our 3\,mm survey (36 discs) are highlighted in red. Discs without a measured stellar mass are not shown in the plot.}
    \label{fig:lupus.sample.3mm.survey}
\end{figure}

The resulting sample encompasses the stellar mass range between 0.1 and 2.8 $\MSun$, with spectral types from M5.5 up to K0. The sample is complete around the Solar-mass range between 0.8 and 1.2\,$\MSun$.
The stellar masses adopted in this paper are reported in \tbref{tb:YSOs}. They were determined from optical-UV spectroscopic measurements by \citet{Alcala:2017aa} and updated by \citet{Alcala:2019aa} to account for the distances drawn from Gaia DR~2 \citep{Gaia-Collaboration:2018aa} (see Table~A.1 in \citealt{Alcala:2019aa}). For all but three sources (Sz~68, RY~Lup, and J16083070-3828268) the stellar masses were derived using pre-main-sequence evolutionary tracks by \citet{Baraffe:2015aa}. For Sz~68, RY~Lup, and J16083070-3828268 we use stellar masses derived using tracks by \citet{Siess:2000qy}.
\figref{fig:lupus.sample.3mm.survey}
summarises the properties of the 3\,mm survey sample in terms of 0.89\,mm integrated flux as a function of stellar mass. The figure shows all the targets of the 0.89\,mm survey \citep{Ansdell:2016qf} except for J15450634-3417378, J16070854-3914075, and J16011549-4152351 for which a stellar mass is not available. 
The figure clearly shows that the 3\,mm survey is 97\% complete above the 0.89\,mm flux median (14.4\,mJy). Overall, the 3\,mm survey sample covers the brightest 55\% of the Class~II discs detected at 0.89\,mm by \citet{Ansdell:2016qf} and is therefore biased towards the most massive discs. 
\section{Observations}
\label{sect:observations}
The ALMA Cycle 4 observations (Project ID: 2016.1.00571.S, PI: M. Tazzari) were performed between 3 and 6 October 2016. For the faintest targets (Sz~123A, SSTc2dJ154508.9-341734, Sz~69, Sz~72, Sz~110) the 2016 observations did not achieve the required sensitivity of 50\microJybeam and were thus observed again on 18 July 2017, achieving a combined sensitivity of 20\muJybeam. The continuum spectral windows were centered on 90.6, 92.5, 102.6 and 104.5 GHz with bandwidths of 1.875 GHz and channel widths of 976.6 kHz. The bandwidth-weighted mean continuum frequency was 97.55 GHz (3.07\,mm). 

Although the spectral setup was optimised to provide optimal continuum sensitivity, it covered the HNC(1-0) spectral line (90.663\,GHz) at low spectral resolution, allowing for a serendipitous detection should the line  have been bright enough. Detailed analysis of the HNC measurement from this dataset is provided in Long et al. (submitted).

The array configuration used forty-three 12m antennas with baselines of 17-3150 m, corresponding to 5.3-1075 k$\lambda$. To optimise the overall time needed for the observations, the correlator was set to integrate for 2 minutes per source on the eighteen brightest targets, 10 minutes per source on the five faintest ones (listed above), and between 4 and 6 minutes for the targets with intermediate brightness, achieving respectively an rms noise of 50, 17, and 30\muJybeam. Data calibration and imaging were performed using CASA 4.7.2 \citep{McMullin:2007aa}. The data was calibrated using the pipeline by ESO and included flux, gain and bandpass calibrations. Flux calibration used observations of J1427-4206, bandpass calibration used observations of J1517-2422, and gain calibration used observations of J1604-4228, J1534-3526, or J1610-3958. We estimated an absolute flux calibration error of 10\% based on variations in the gain calibrators. 

We successfully performed self-calibration on the four brightest sources in the sample (Sz~68, Sz~82, Sz~83, Sz~98). In all cases we performed two phase-only self-calibration steps: in the first step we sought solutions across the whole scan length, and in the second step over 60 seconds. We did not find any appreciable improvement in the noise and signal-to-noise properties for phase-only self-calibration steps with smaller time intervals or amplitude self-calibration. In all cases we improved rms noise by 20\% and a signal-to-noise ratio by 60\%. 

\section{Results}
\label{sect:results}

\subsection{Continuum emission at 3\,mm}
\label{sect:results.3mmcontinuum}
\begin{figure*}
\includegraphics[width=\hsize]{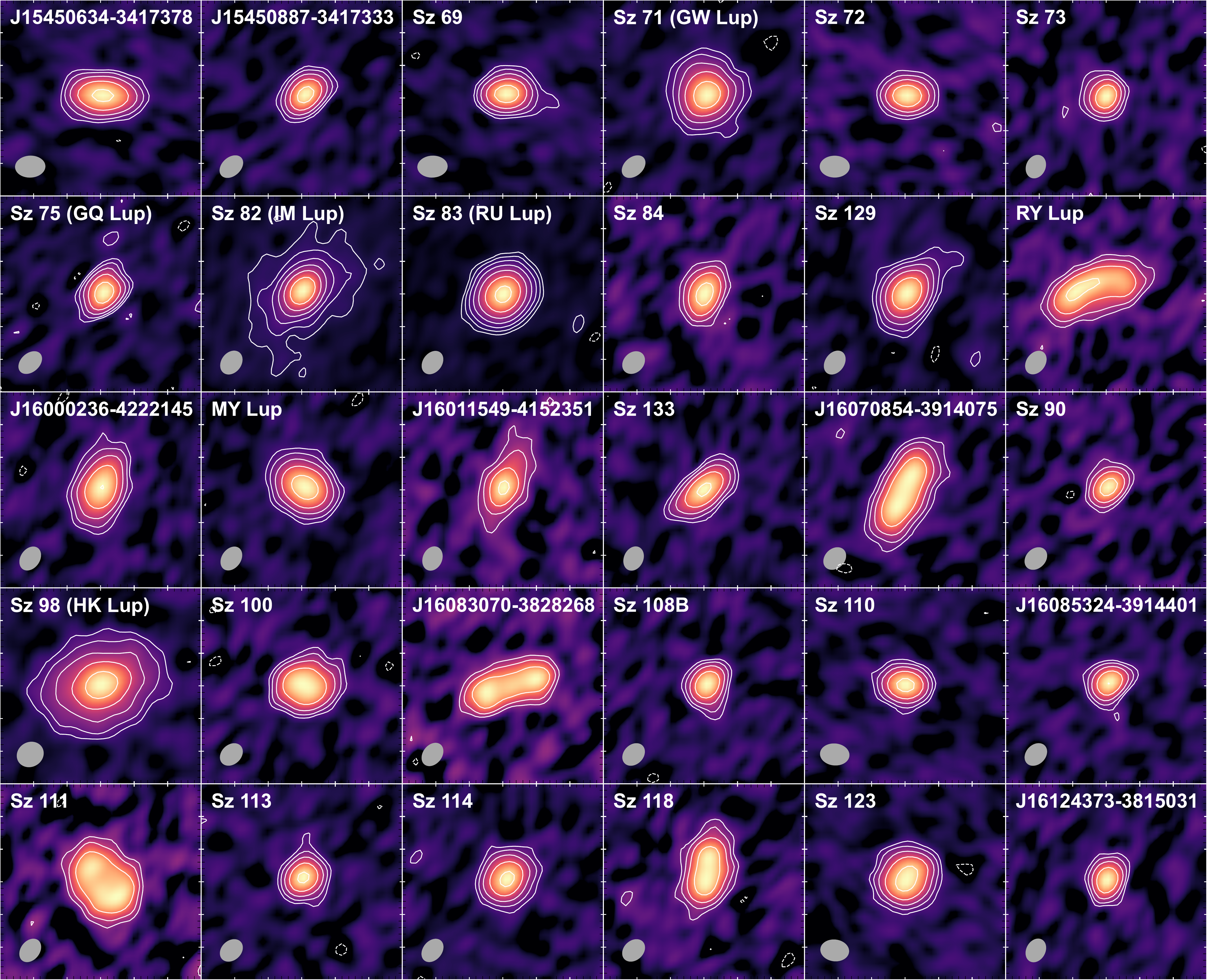}
\caption{Gallery of the 3\,mm continuum emission around single stars, sorted as in Table~\ref{tb:YSOs}. Each image covers a field of view of $3\arcsec\times 3\arcsec$. The synthesized beam (Briggs weighting, robust 0.5) is represented as a grey ellipse. The colour scale is different in each panel, normalised to the brightness peak. Contours are drawn at -3, 3, 6, 12, 48, 96, 192 times the rms noise in each image.}
\label{fig:gallery.continuum.singles}
\end{figure*}

\begin{figure*}
\includegraphics[width=\hsize]{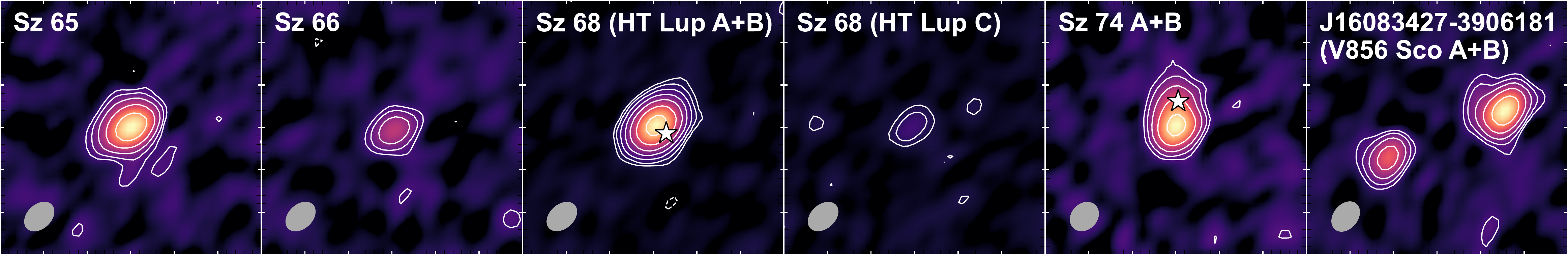}
\caption{Gallery of the 3\,mm continuum emission around binaries and the triple system Sz~68, sorted as in Table~\ref{tb:YSOs}. 
Sz~65 and Sz~66 are separated by $6.4\arcsec$, HT~Lup A and B are separated by 0.16$\arcsec$, and in turn they are distant 2.8$\arcsec$ from HT~Lup C, Sz~74 A and B are separated by 0.3$\arcsec$.
The location of HT~Lup B (separation 0.16$\arcsec$ and Sz~74 B is indicated with a white star.
Each image covers a field of view of $3\arcsec\times 3\arcsec$. The synthesized beam (Briggs weighting, robust 0.5) is represented as a grey ellipse. The colour scale is the same in within each system, normalised to the brightness peak. Contours are drawn at -3, 3, 6, 12, 48, 96, 192 times the rms noise in each image. }
\label{fig:gallery.continuum.binaries}
\end{figure*}

Figures~\ref{fig:gallery.continuum.singles} and \ref{fig:gallery.continuum.binaries} present the continuum images synthesised from the calibrated visibilities. To produce the images we apply the CASA \texttt{tclean} task to the full dataset before any channel averaging. Imaging was performed using the Briggs weighting scheme (robust parameter 0.5), which yields an optimal combination of resolution and sensitivity. We used the multi-frequency synthesis deconvolver with multiple spatial scales set to a point source, the size of the synthesised beam, and 3 times the synthesised beam in order to improve image fidelity. Given the large fractional bandwidth of the data (ratio of effective bandwidth to central average frequency, $\approx8$\%), we performed a multi-term clean (\texttt{nterms}$=2$), which better accounts for spectral index variations across the image.
The average synthesized beam size is $0.39\arcsec\times0.29\arcsec$, corresponding to approximately 60 $\times$ 46 au at the average distance of Lupus discs (160\,pc; \citealt{Manara:2018aa}). 

When observed at 3\,mm, most discs appear regularly shaped. In a few cases, they exhibit an unusual shape owing to a peculiar morphology or viewing geometry. Discs around J16070854-3914075 and J16083070-3838268 are seen almost edge-on, thus appearing particularly elongated. In addition, they are also transition discs \citep{van-der-Marel:2018aa}. RY~Lup is genuinely side-lobed (see also the 1.3\,mm observations in \citealt{Ansdell:2018aa}) and is observed at a high inclination. Finally, transition discs around Sz~84, Sz~100, Sz~111, Sz~118, and Sz~123 also exhibit some uneven structure. MY~Lup is the only transition disc in the sample that does not show a clear cavity, however this is likely due to the high inclination (larger than 70$^\circ$; \citealt{Tazzari:2017aa}) and the limited resolution of these observations.

In Table~\ref{tb:measurements.band3} we report the integrated flux measured at 3\,mm and the synthesized beam size and position angle. We detect all the sources except Sz~91, which was observed with an erroneous (too large) rms noise requirement due to a typo in the observational setup. From \citet{Canovas:2016aa} observations at 2.7\,mm we expect Sz~91 to be spatially resolved at the resolution of these new 3\,mm observations ($\sim0.35\arcsec$), thus it is not possible to set a simple upper limit on its 3\,mm flux from this data.

The reported integrated fluxes are measured from continuum images synthesized with natural weighting (in order to maximise signal-to-noise) using circular aperture photometry. The aperture radius for each source is determined by a curve-of-growth method in which a circular aperture of increasing radius is applied until the enclosed flux becomes constant at a 3$\sigma$ level. We find that typically the aperture radius at which the enclosed flux plateaus tightly encloses the 3$\sigma$ contours, indicating that the contribution to the flux from the outer disc regions emitting just below the sensitivity of the observations is minimal. The uncertainty on the total flux is measured as the standard deviation of the flux measured within 10 apertures of same size in a region of the field of view far from the source and with no other emission. The uncertainty on the total flux reported in Table~\ref{tb:measurements.band3} does not include the systematic 10\% absolute flux uncertainty. We note that \citet{Ansdell:2016qf} used a different method to measure the integrated fluxes: they did a two-dimensional Gaussian parametric fit of the visibilities with the \texttt{uvmodelfit} CASA task. As a check, we have used the same method on our 3\,mm observations and found integrated fluxes that agree extremely well with those measured with the circular aperture method. 
\begin{table*}
\caption{ALMA continuum measurements at 3\,mm, inferred dust masses, and 1-3\,mm spectral indices. }
\begin{tabular}{llcrcrrr}
\midrule
\toprule
Name & Other Name & \multicolumn{2}{c}{Beam} & rms & $F_\mathrm{3\,mm}$  & $M_\mathrm{dust}$   &   $\alphamm$\\
\cline{3-4} \\[-0.3cm]
 & & Size & P.A. & & & & \\
  &   &  ($\arcsec \times \arcsec$) & ($^{\circ}$)  & ($\mu$Jy beam$^{-1}$)&   (mJy)               &  ($M_\oplus$)  &                \\   
   \midrule
Sz 65                    &             &    $ 0.46\times0.35 $&  -44&      53 & $  5.45^{\pm0.17}$& $ 14.48^{\pm0.44}$& $  1.96^{\pm0.11}$ \\ 
Sz 66                    &             &    $ 0.46\times0.35 $&  -44&     53 &  $  1.21^{\pm0.14}$& $  3.31^{\pm0.37}$& $  1.98^{\pm0.14}$ \\ 
J15450634-3417378        &             &    $ 0.53\times0.39 $&   88&      20 & $  1.03^{\pm0.07}$& $  3.41^{\pm0.22}$& $  2.12^{\pm0.12}$ \\ 
J15450887-3417333        &             &    $ 0.46\times0.35 $&  -46&      47 & $  3.86^{\pm0.16}$& $ 10.27^{\pm0.42}$& $  1.96^{\pm0.12}$ \\ 
Sz 68                    &      HT Lup &    $ 0.46\times0.35 $&  -46&      55 & $ 13.74^{\pm0.14}$& $ 36.03^{\pm0.37}$& $  1.89^{\pm0.11}$ \\ 
Sz 69                    &             &    $ 0.53\times0.39 $&   88&      18 & $  1.64^{\pm0.06}$& $  4.32^{\pm0.17}$& $  1.85^{\pm0.12}$ \\ 
Sz 71                    &      GW Lup &    $ 0.46\times0.35 $&  -47&      46 & $  9.82^{\pm0.21}$& $ 35.41^{\pm0.76}$& $  2.24^{\pm0.11}$ \\ 
Sz 72                    &             &    $ 0.53\times0.39 $&   88&      18 & $  0.95^{\pm0.06}$& $  3.01^{\pm0.18}$& $  2.13^{\pm0.12}$ \\ 
Sz 73                    &             &    $ 0.43\times0.36 $&  -23&      30 & $  1.97^{\pm0.08}$& $  6.57^{\pm0.28}$& $  2.16^{\pm0.12}$ \\ 
Sz 74                    &             &    $ 0.44\times0.38 $&  -39&      24 & $  2.55^{\pm0.11}$& $  6.38^{\pm0.26}$& $  1.67^{\pm0.12}$ \\ 
Sz 75                    &      GQ Lup &    $ 0.46\times0.35 $&  -45&      46 & $  4.60^{\pm0.11}$& $ 15.75^{\pm0.37}$& $  2.24^{\pm0.11}$ \\ 
Sz 82                    &      IM Lup &    $ 0.45\times0.35 $&  -33&      54 & $ 20.45^{\pm0.51}$& $131.18^{\pm3.27}$& $  2.67^{\pm0.16}$ \\ 
Sz 83                    &      RU Lup &    $ 0.45\times0.36 $&  -34&      65 & $ 22.74^{\pm0.31}$& $ 95.30^{\pm1.30}$& $  2.32^{\pm0.11}$ \\ 
Sz 84                    &             &    $ 0.45\times0.38 $&  -41&      24 & $  1.59^{\pm0.04}$& $  6.66^{\pm0.16}$& $  2.39^{\pm0.11}$ \\ 
Sz 129                   &             &    $ 0.46\times0.35 $&  -32&      47 & $  9.90^{\pm0.17}$& $ 41.47^{\pm0.70}$& $  2.30^{\pm0.11}$ \\ 
J15592838-4021513        &      RY Lup &    $ 0.45\times0.35 $&  -32&      43 & $  7.18^{\pm0.26}$& $ 60.56^{\pm2.16}$& $  2.89^{\pm0.12}$ \\ 
J16000236-4222145        &             &    $ 0.46\times0.36 $&  -31&      43 & $  6.83^{\pm0.18}$& $ 28.32^{\pm0.75}$& $  2.27^{\pm0.11}$ \\ 
J16004452-4155310        &      MY Lup &    $ 0.45\times0.35 $&  -32&      48 & $  8.83^{\pm0.20}$& $ 37.96^{\pm0.84}$& $  2.37^{\pm0.11}$ \\ 
J16011549-4152351        &             &    $ 0.45\times0.36 $&  -21&      29 & $  2.57^{\pm0.13}$& $ 18.45^{\pm0.93}$& $  2.74^{\pm0.12}$ \\ 
Sz 133                   &             &    $ 0.45\times0.36 $&  -23&      30 & $  3.96^{\pm0.08}$& $ 14.60^{\pm0.30}$& $  2.26^{\pm0.11}$ \\ 
J16070854-3914075        &             &    $ 0.45\times0.38 $&  -42&      24 & $  5.26^{\pm0.12}$& $ 25.44^{\pm0.56}$& $  2.26^{\pm0.11}$ \\ 
Sz 90                    &             &    $ 0.45\times0.38 $&  -42&      22 & $  1.17^{\pm0.06}$& $  4.92^{\pm0.26}$& $  2.31^{\pm0.12}$ \\ 
Sz 91                    &             &    $ 0.45\times0.35 $&  -35&      43 & $-$                 &   $-$                 &   $-$                  \\ 
Sz 98                    &      HK Lup &    $ 0.45\times0.36 $&  -36&      43 & $ 12.73^{\pm0.25}$& $ 50.73^{\pm1.01}$& $  2.31^{\pm0.11}$ \\ 
Sz 100                   &             &    $ 0.45\times0.38 $&  -43&      23 & $  3.23^{\pm0.09}$& $  9.05^{\pm0.26}$& $  2.24^{\pm0.11}$ \\ 
J16083070-3828268        &             &    $ 0.45\times0.35 $&  -35&      41 & $  4.77^{\pm0.29}$& $ 28.74^{\pm1.72}$& $  2.65^{\pm0.12}$ \\ 
J16083427-3906181        &    V856 Sco &    $ 0.45\times0.35 $&  -36&      45 & $  5.14^{\pm0.16}$& $ 14.71^{\pm0.45}$& $  1.89^{\pm0.11}$ \\ 
Sz 108B                  &             &    $ 0.45\times0.38 $&  -41&      23 & $  1.75^{\pm0.07}$& $  6.72^{\pm0.28}$& $  2.16^{\pm0.12}$ \\ 
Sz 110                   &             &    $ 0.52\times0.40 $&   83&      19 & $  1.32^{\pm0.04}$& $  3.71^{\pm0.12}$& $  1.94^{\pm0.12}$ \\ 
J16085324-3914401        &             &    $ 0.45\times0.38 $&  -41&      22 & $  1.49^{\pm0.07}$& $  4.85^{\pm0.22}$& $  2.04^{\pm0.12}$ \\ 
Sz 111                   &             &    $ 0.45\times0.35 $&  -34&      44 & $  5.83^{\pm0.17}$& $ 39.15^{\pm1.11}$& $  2.71^{\pm0.11}$ \\ 
Sz 113                   &             &    $ 0.45\times0.38 $&  -44&      23 & $  2.07^{\pm0.06}$& $  6.07^{\pm0.19}$& $  1.88^{\pm0.11}$ \\ 
Sz 114                   &             &    $ 0.45\times0.35 $&  -36&      44 & $  5.60^{\pm0.15}$& $ 22.26^{\pm0.59}$& $  2.25^{\pm0.11}$ \\ 
Sz 118                   &             &    $ 0.45\times0.38 $&  -43&      23 & $  3.11^{\pm0.10}$& $ 14.88^{\pm0.49}$& $  2.39^{\pm0.12}$ \\ 
Sz 123                   &             &    $ 0.51\times0.39 $&   81&      19 & $  2.53^{\pm0.04}$& $  9.46^{\pm0.15}$& $  2.20^{\pm0.11}$ \\ 
J16124373-3815031        &             &    $ 0.44\times0.36 $&  -26&      30 & $  1.84^{\pm0.07}$& $  6.70^{\pm0.25}$& $  2.20^{\pm0.12}$ \\ 
\bottomrule
\end{tabular}
\begin{flushleft}
\textbf{Note.}
Name is the designation used in this paper (with notable alternative names where available).
Beam is the FWHM of the synthesized beam obtained with natural weighting, rms is the image noise, $F_\mathrm{3\,mm}$ is the integrated flux at 3\,mm measured with a curve of growth method on circular aperture photometry. 
The 0.89\,mm fluxes used to compute $\alphamm$ are from \citet{Ansdell:2016qf} except for Sz~75 for which we used the flux measured by \citet{MacGregor:2017aa}. Note that 7 discs (Sz~68, Sz~71, Sz~82, Sz~83, Sz~114, Sz~129, and MY~Lup) have been targeted by the DSHARP survey \citep{Andrews:2018aa}. A machine-readable version of this table is available online (see the Data Availability statement).
\end{flushleft}
\label{tb:measurements.band3}
\end{table*}
\subsection{The 1-3\,mm spectral indices}
\label{sect:results.alpha}
We now compute the disc-integrated spectral indices between 0.89 and 3\,mm, deferring to a forthcoming paper \citep{Tazzari:2020ab} the study of their spatial variations.
The integrated spectral indices $\alphamm$ have been computed using the 0.89\,mm fluxes measured by \citet{Ansdell:2016qf} and assuming a linear spectral slope between 0.89 and 3.1\,mm: $F_\nu\propto \nu^{\alphamm}$.
The uncertainties on the spectral indices are propagated from the flux uncertainties at the two wavelengths and include a 10\% flux calibration uncertainty at both wavelengths. In a few cases we obtain values $\alphamm<2.0$ that are however still compatible at a 2$\sigma$ level with $\alphamm=2$. In only one case (Sz~74) $\alphamm<2$ at a 3$\sigma$ level. A measured value of $\alphamm<2$ could be due to the contamination from non-thermal emission processes, due to the departure of thermal emission from the Rayleigh-Jeans regime, or due to extremely high optical depth. We can exclude the first scenario since the non-thermal contribution needed at 3\,mm to obtain such a low spectral index would be much larger than the typical measured upper limits \citep{Ubach:2012ul}. We can also exclude the second scenario as these low spectral indices below 2 are obtained for discs that are very small, in which the temperature is likely to be high enough to ensure the radiation is emitted in Rayleigh Jeans regime. 
It is more likely that in the case of Sz~74, a close binary, the low spectral index reflects the highly optically thick contribution from the primary disc, which is expected to be truncated due to tidal interaction.

Some of the targets in our sample were observed at millimetre wavelengths by \citet{Lommen:2007aa} and \citet{Ubach:2012ul} using the Australia Telescope Compact Array (ATCA) at lower resolution and more limited sensitivity. 
Both these studies targeted Sz~68, Sz~71, and Sz~98; \citet{Ubach:2012ul} targeted also Sz~65, Sz~66, Sz~75, RY~Lup, Sz~111 and \citet{Lommen:2007aa} targeted also Sz~82 and Sz~83. These studies obtained observations at 1-3$\arcsec$ resolution and a sensitivity of 0.5-2\mJybeam, measuring 3.3\,mm fluxes that are compatible with those presented in this paper except for Sz~82, Sz~83, RY~Lup, and Sz~98, for which we obtain larger 3\,mm fluxes and thus smaller $\alphamm$ spectral indices. 
\citet{Lommen:2007aa} and \citet{Ubach:2012ul}, however, do not report the total uncertainty on their measured integrated 3\,mm flux: an inspection of their deprojected visibility profiles reveals that the data is indeed very noisy, and for Sz~82, RY~Lup, and Sz~98 the short baseline visibilities appear compatible with the integrated flux that we measure from our observations.

\begin{figure}
\centering
\includegraphics[scale=1]{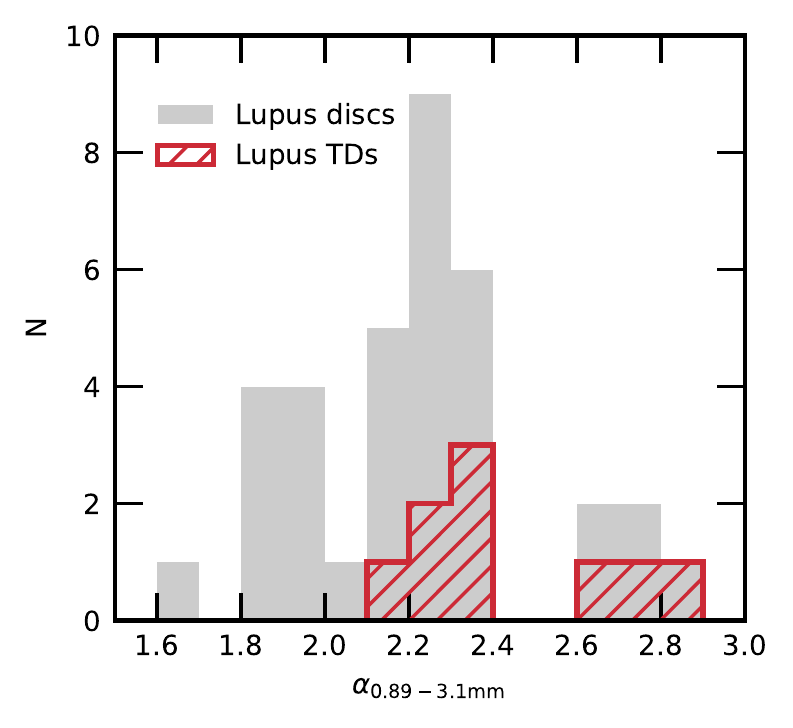}
\caption{Histogram of the 1-3\,mm spectral index measurements in the Lupus sample. Lupus transition discs (TDs) as defined in \citet{Ansdell:2018aa} are highlighted in red.}
\label{fig:alpha.hist}
\end{figure}

\figref{fig:alpha.hist} presents the histogram of the spectral index measurements, which clearly highlights that resolved transition discs (TDs) in Lupus defined as in \citet{Ansdell:2018aa} (see also \citealt{van-der-Marel:2018aa}) tend to have a higher spectral index compared to the bulk of the disc population. 
The same trend was observed in the Lupus discs survey by \citet{Ansdell:2018aa} in the 0.89-1.3\,mm wavelength range and by \cite{Pinilla:2014aa} in a more heterogeneous set of observations at 0.89 and 3\,mm of bright discs in the Taurus, Ophiuchus, and Orion star-forming regions. 
\cite{Pinilla:2014aa} reported average spectral indices of $\alpha_\mathrm{TD}=2.7\pm0.1$ for transition discs and $\alpha_\mathrm{PPD}=2.2\pm0.1$ for protoplanetary discs (PPD) with no known cavities. In this study we find similar values of $\alpha_\mathrm{TD}=2.5\pm0.1$ and $\alpha_\mathrm{PPD}=2.14\pm0.06$, respectively.
For completeness we note that \citet{Ansdell:2018aa} considered a complete survey of the Lupus Class~II discs, out of which this study selected the bright end of the population. The sample used by \citet{Pinilla:2014aa} is also biased towards the brightest discs but encompasses discs from different regions and is much more dishomogeneous in terms of observational setups.
Evaluating the universality of this trend would require a detailed physical modelling of these sources and a careful consideration of the observational biases, which is beyond the scope of this paper.

There are three binaries (Sz~65+Sz~66, Sz~74, and V856~Sco/J16083427-3906181) and one triple system (Sz~68/HT Lup) in our sample. They have very different separations: $6.4\arcsec$ (960\,au) for Sz~65+Sz~66, $1.45\arcsec$ (225\,au) for V856 Sco \citep{Ansdell:2018aa}, and $0.3\arcsec$ (45\,au) for Sz~74 \citep{Ansdell:2018aa}. We note that Sz~65+Sz~66 and V856~Sco have both components individually detected at all ALMA Bands, while for Sz~74 the angular resolution gives us only tentative evidence of the detection of the secondary component (see Fig.~\ref{fig:gallery.continuum.binaries}). 
Sz~68 is a triple system with A+B components (not resolved individually) at $0.14\arcsec$ (20\,au) separation \citep{Kurtovic:2018ur}, which in turn are distant $2.8\arcsec$ ($\sim450\,$au) from the C component (detected at  11$\sigma$).
For Sz~74 and Sz~68 \tbref{tb:measurements.band3} reports the integrated flux and the spectral index of their A+B components.

\subsection{Comparison to other regions}
\begin{figure*}
\centering
\includegraphics[scale=1]{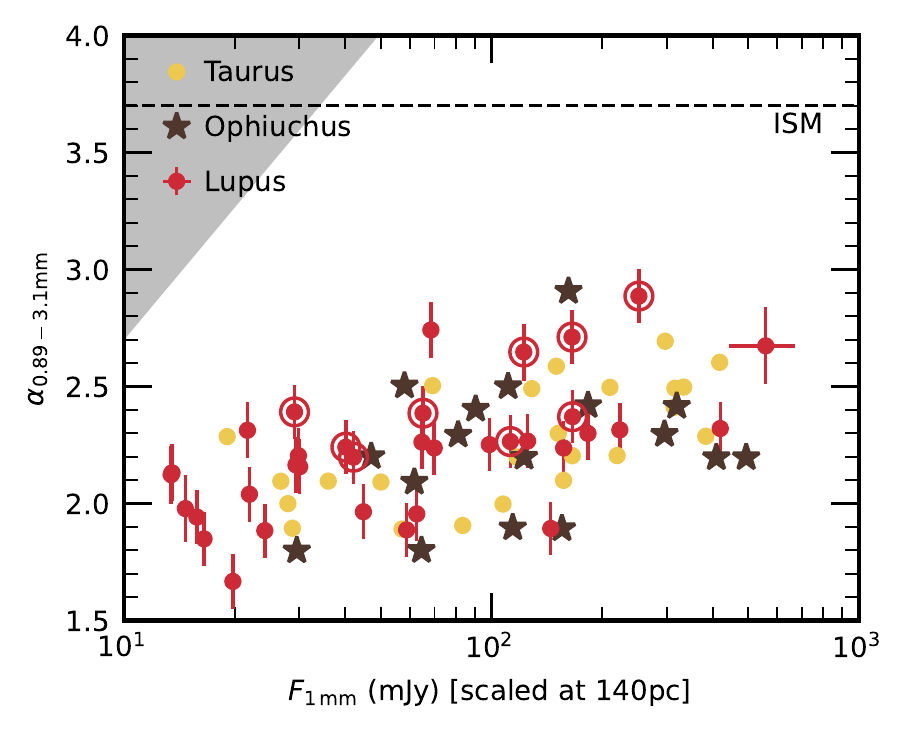}
\hfill
\includegraphics[scale=1]{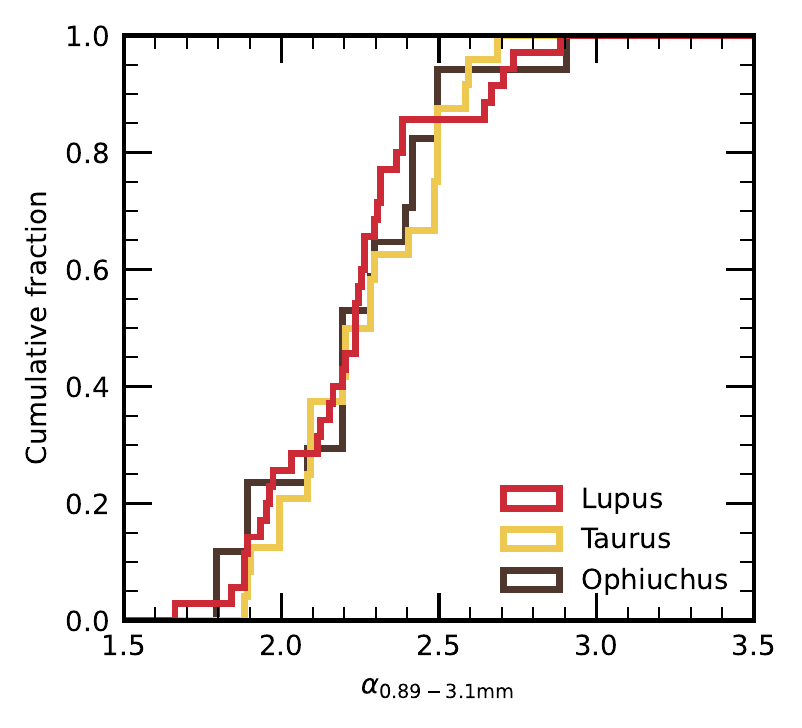}
\caption{\textit{(Left)} Spectral index between 1 and 3\,mm as a function of integrated 1\,mm flux for Lupus discs (red), Ophiuchus (dark brown), and Taurus (yellow). Lupus TDs are marked with an additional red circle. 
The dark shaded region represents the sensitivity cut-off of our ALMA observations, i.e. where observations are \textit{not} sensitive anymore. The dashed line is the typical $\alphamm$ of the optically thin emission of ISM dust. 
\textit{(Right)} Normalised cumulative distribution of the spectral index measurements shown in the left panel.}
\label{fig:alpha.vs.flux.regions}
\end{figure*}

In Figure~\ref{fig:alpha.vs.flux.regions} we compare the spectral indices of the Lupus discs with those of other nearby star-forming regions like Taurus and Ophiuchus as a function of their integrated 1\,mm flux (scaled at a distance of 140\,pc). The spectral indices in Taurus and Ophiuchus are taken from a compilation by \citet{Ricci:2010eu,Ricci:2010fv} who fitted the spectral energy distribution in the 0.89-3.0\,mm range for a sample of Class~II sources with known stellar properties, without envelope contamination, and with no evidence of companion at 10-400\,au scale  (for more details, we refer to their papers). The star-forming regions chosen for this comparison are all relatively young, with mean ages between 1 and 3\,Myr \citealt{Alcala:2017aa}).
Compared to the spectral index $\alpha_\mathrm{ISM}\simeq 3.7$ typical of the optically thin emission of interstellar medium grains, all these regions exhibit much lower spectral indices, the bulk of the measurements lying between 1.8 and 2.5. Moreover, the similarity of the spectral indices distribution among the different regions, as highlighted by the cumulative fractions plotted in the right panel of Fig.~\ref{fig:alpha.vs.flux.regions}, is particularly striking. Two population Anderson-Darling tests applied to the spectral index distribution of these three regions confirm that they are indistinguishable in a statistically significant way ($p$-value$>0.25$ for the null hypothesis of spectral index measurements being drawn from the same parent distribution).

It is worth noting that the lack of discs with low sub-millimeter (or millimeter) flux and high spectral indices could in principle reflect an observational bias induced by the sensitivity limitations at the longest observing wavelength, 3\,mm in this case. To evaluate whether we are affected by this issue, in Figure~\ref{fig:alpha.vs.flux.regions} we plot the sensitivity cut imposed by the combined sensitivity of the observations presented in this paper at 3\,mm and those by \citet{Ansdell:2016qf} at 1.3\,mm. Even for the smallest 1.3\,mm fluxes, the sensitivity cut is comfortably far from the maximum spectral index inferred from our measurements and we can safely assume that the lack of discs in such region is genuine. 
Provided that the sample of Lupus discs imaged at 3\,mm is complete for discs brighter than the 0.89\,mm flux median (14.4\,mJy), we can exclude that other Lupus discs can populate such region of the plot. 
Although it may seem that Lupus discs are characterised by a weak positive correlation between $\alphamm$ and $F_\mathrm{1\,mm}$, this is mostly driven by the three TDs with large spectral index and large millimeter flux. Excluding TDs from the sample no statistically significant correlation can be found. Given the low number statistics (3 discs) driving the possible positive correlation, we do not explore this further.

\section{Discussion}
\label{sect:discussion}
Spectral indices at sub-millimeter and millimeter wavelengths provide us with valuable information on the optical properties of the population of large grains residing in the discs midplane. 
In Sect.~\ref{sect:discussion.grain.growth} we discuss the constraints that our new 3\,mm measurements set on the average grain properties in the Lupus discs assuming that most of the observed emission is optically thin.
In Sect.~\ref{sect:discussion.size.luminosity} we discuss the millimeter continuum size-luminosity relation \citep{Tripathi:2017aa,Andrews:2018ab} in light of new constraints posed by the 3\,mm observations presented here.

\subsection{Implications for grain growth}
\label{sect:discussion.grain.growth}
Thanks to their sensitivity to the thermal emission of dust grains harboured in the dense and cold disc midplane, sub-millimeter and millimeter observations constitute a powerful probe of the early phase of grain growth from sub-micron to mm sizes \citep[][, and references therein]{Testi:2014kx}. At sub-millimeter and millimeter wavelengths the dust emission is mostly optically thin and the slope $\alpha_{\mathrm{mm}}$ of the (sub-)mm SED (namely, the spectral index)
\begin{equation}
    \alpha_\mathrm{mm} = \frac{\mathrm{d}\log F_\nu}{\mathrm{d}\log\nu}
\end{equation}
can be related in first approximation to the dust opacity power-law slope $\beta$, being the opacity $\kappa_\nu \propto \nu^\beta$, as
\begin{equation}
\label{eq:def.beta.opacity}
    \alpha_\mathrm{mm}\approx\beta + 2\,,
\end{equation}
where the further assumption that the radiation is emitted in Rayleigh-Jeans regime was made. 
If we consider a power-law grain size distribution $n(a)\propto a^{-q}$ for $a_\mathrm{min}\leq a \leq a_\mathrm{max}$ ($a$ being the radius of the emitting grain), it is possible to show \citep{Miyake:1993lr} that the dust power-law index $\beta$  depends strongly on $a_\mathrm{max}$ (provided that $a_\mathrm{min}<1\mu $m). 
Typical  values of $\alpha_{\mathrm{1-3mm}}\sim 3.7$ are found for the relatively small ISM dust grains ($a_\mathrm{max}\approx 0.25\mu$m),
while \citet{Natta:2004yu} showed that dust grains with sizes $\beta \leq 1$ ($\alpha_\mathrm{1-3mm}\leq 3$) can be safely interpreted as evidence of large grains ($a_\mathrm{max}\ge 1$mm) for a wide range of dust composition, porosity and size distribution (see also \citealt{Draine:2006ty}).

Using the new 3\,mm data presented here, we find that the mean spectral index in the Lupus discs is $\alphamm=2.23 \pm 0.06$, which corresponds to a nominal average dust opacity $\beta= 0.23 \pm 0.06$ according to Eq.~\eqref{eq:def.beta.opacity}.
By comparing this result with theoretical dust opacity models based on Mie theory and a range of compositions \citep[e.g.,][]{Draine:2006ty,Birnstiel:2018aa}, we find that $\beta \sim 0.2-0.5$ requires a grain populations dominated by large grains ($q\leq3$) with maximum grain size at least larger than 1\,mm. 

Global dust evolution models \citep{Brauer:2007aa,Brauer:2008kq} predict very short lifetimes for large grains: as soon as they grow past mm sizes at large distances from the star, they are expected to undergo rapid inward migration due to the loss of angular momentum to the gaseous component of the disc \citep{Weidenschilling:1977lr}.
Observationally, the loss of large grains is expected to make discs evolve very quickly towards large values $\alphamm> 3.5$ \citep{Birnstiel:2010aa}. However, the observational evidence of low values of $\alphamm\simeq2.23$ (hence, $\beta\simeq0.23$ and $\amax>1\,$mm) gathered not only in the Lupus region, but also in the coeval (1-3\,Myr old) Taurus and Ophiuchus regions  (cf. Fig.~\ref{fig:alpha.vs.flux.regions}),  can be interpreted in terms of a high occurrence of dust retention mechanisms. Pressure maxima created by strong gas inhomogeinities \citep{Pinilla:2012xy} or by a forming planet embedded in the disc \citep{Whipple:1972tk,Brauer:2008kq,Pinilla:2012aa}, streaming instability \citep{Youdin:2005aa}, and dust accumulation in vortices \citep{1995A&A...295L...1B,Klahr:1997fk,Lyra:2009aa,Barge:2017aa} are all effects that can potentially slow down or even completely halt the drift of large grains. In this work we focused on the spatially-integrated analysis of these new 3\,mm observations at moderate resolution, which do not allow us to infer which of these mechanisms are shaping the Lupus discs. 
However, 7 discs in our sample have been targeted by ALMA at 1.3\,mm and extreme angular resolution by the DSHARP \citep{Andrews:2018aa} survey (see note in \tbref{tb:YSOs}): while two discs (Sz~114 and MY~Lup) exhibit a rather smooth surface brightness, five of them (Sz~68, Sz~71, Sz~82, Sz~83, and Sz~129) appear as highly structured, lending support to the scenario in which the large grains are mostly concentrated in long-lived disc structures. 

A further argument supporting the presence of large grains is that even TDs have relatively small $\alphamm\simeq 2.5$ values. In these cavity-bearing discs, the spectral index is determined by the emission of the outer disc, i.e. mostly free from the contamination of a possibly optically thick inner disc. Even considering the caveat that TDs might still have very small inner discs and (or) very narrow outer rings, the fact that they exhibit $\alphamm\simeq 2.5$ suggests that optical depth is genuinely small and the inferred face-value $\beta\simeq0.5$ can be robustly interpreted with the presence of large grains.

\subsubsection{Caveats}
A natural caveat of interpreting the low spectral index values in terms of mm-sized grains is that an increased optical depth or any deviations from Rayleigh-Jeans regime would naturally alter the simple relation in Eq.~\eqref{eq:def.beta.opacity} between $\alphamm$ and $\beta$, with both effects tending to bias $\alphamm$ towards lower values \citep[e.g., see the discussion in][]{Huang:2018aa}.  \citet{Tazzari:2017aa} analysed the 0.89\,mm ALMA observations of most of the discs in this sample with a two-layer disc model \citep{1997ApJ...490..368C,Dullemond:2004ly} finding that most discs should be  emitting in the Rayleigh-Jeans regime. However, we cannot exclude that for large discs, the emission from the outer and colder parts of the disc with temperatures approaching 7\,K could contribute a non negligible amount of emission. We note that the largest disc in our sample, Sz~82/IM~Lup, has a relatively large spectral index $\alphamm=2.67\pm0.16$, suggesting that the contribution of non-Rayleigh-Jeans emission to its observed flux is small.

It is worth highlighting that in this analysis we have considered only the absorption component of dust opacity. Although at low optical depths this is essentially accurate, if the optical depth is high and large grains are present, dust scattering is expected to be an important contributor to the total millimeter opacity \citep{Liu:2019aa,Zhu:2019aa}. Scattering acts by reducing the dust emission below its nominal black-body value (with a spectral dependency of this effect that in turn depends on that of the albedo), and it can potentially change the nominal $\alphamm\sim2.0$ of an optically thick disc in Rayleigh-Jeans regime to a value as low as 1.6 or as high as 2.5 \citep{Carrasco-Gonzalez:2019aa}.
The disc-integrated analysis that we performed in this study does not allow us to robustly quantify the optical depth in these discs and we cannot therefore rule out this alternative explanation for the low spectral indices measured in these Lupus discs. It is worth mentioning that if optical depth will turn out to be small, then the above results based on absorption remain valid. A forthcoming spatially-resolved study of the multi-wavelength observations gathered for the Lupus discs will allow us to put firmer constraints on the radial variations of the optical depth, on its dependency on frequency, and therefore on the dust properties in the Lupus discs \citep{Tazzari:2020ab}.

A third caveat is represented by the observational sample (Sect.~\ref{sect:sample}), which includes only the brightest Lupus discs. Being the sub-millimeter and millimeter flux a rough proxy for the mass of the emitting dust, the sample is clearly biased towards the most massive discs, which can also be preferentially characterised by enhanced optical depth at small radii w.r.t. the bulk disc population. Extending the present 3\,mm survey to the fainter disc population is the necessary next step to quantify this potential bias.

\subsection{Disc dust masses}
At millimeter wavelengths the disc emission is typically optically thin, allowing us to infer the disc dust mass from the integrated flux given some assumptions on the opacity and the temperature of the emitting dust \citep{Beckwith:1990qf}. Despite admittedly simplified, this approach has been extensively used to infer disc masses from sub-millimeter and millimeter photometric surveys \citep[see, e.g.,][]{Andrews:2005fk,Andrews:2013qy}, and more recently from ALMA continuum surveys \citep{Ansdell:2016qf,Ansdell:2017aa,Pascucci:2016aa,Barenfeld:2016lr,Long:2018aa,Cieza:2019aa}.

We compute disc dust masses ($M_\mathrm{dust}$) from the measured 3\,mm integrated fluxes as \citet{Hildebrand:1983fk}:
\begin{equation}
\label{eq:dust.mass.from.flux}
    M_\mathrm{dust} = \frac{F_\nu\, d^2}{\kappa_\nu B_\nu(T_\mathrm{dust})} 
\end{equation}
where $d$ is the distance inferred from Gaia DR2 measurements \citep{Bailer-Jones:2018aa}, $\kappa_\nu$ is the dust opacity and $T_\mathrm{dust}$ is the dust temperature. To ease the comparison with other measurements in literature (especially with the previous Lupus surveys at 0.89 and 1.3\,mm), we adopt a power-law opacity  $\kappa_\nu=3.37\,(\nu/337\,\mathrm{GHz})^\beta\,\mathrm{cm}^2 \mathrm{g}^{-1}$, with $\beta$ measured for each disc as $\beta=\alphamm-2$ and a normalisation such that the opacity at 0.89\,mm (337\,GHz) matches that used by \citet{Ansdell:2016qf}.
We further assume that the dust is isothermal, with $T_\mathrm{dust}=20\,$K. 
The latter assumption is justified since the bulk of the mm emission is expected to originate in the nearly isothermal outer disc regions. 

The dust masses that we infer from the 3\,mm fluxes are reported in \tbref{tb:measurements.band3}. In order to reduce the biases in comparing these masses with those inferred from previous 0.89\,mm \citep{Ansdell:2016qf} and 1.3\,mm \citep{Ansdell:2018aa} observations of the same Lupus discs, we re-compute the latter ones using Eq.~\eqref{eq:dust.mass.from.flux} and up-to-date distances computed using Gaia DR2 measurements \citep{Bailer-Jones:2018aa}. 
In general, the masses inferred from the 3\,mm fluxes are smaller than those inferred from the 0.89 and 1.3\,mm ones. Median $M_\mathrm{dust}$ values for this sample of Lupus discs inferred at 3, 1.3, and 0.89\,mm are $22\pm2$, $24\pm2$, and $27\pm2$ Earth masses, respectively. Under the assumption of perfectly optically thin emission, this result would imply that the grains contributing to the 3\,mm emission are on average $\sim20\%$ less abundant than the (roughly 3 times) smaller grains emitting at 0.89\,mm. We shall note, however, that these dust masses are highly sensitive to the dust opacity spectral index $\beta$ (i.e., on $\alphamm$): if some of the emission is optically thick (e.g., emission arising from narrow rings unresolved in the current observations), the $\beta$ value to be used in Eq.~\ref{eq:dust.mass.from.flux} would be larger, which would result in a smaller opacity at 3\,mm and therefore in larger dust masses. Multi-wavelength observations at matching high angular resolution are needed in order to resolve the disc structure and break this degeneracy and improve the estimates on the dust mass.

\subsection{Millimeter continuum size-luminosity relation: new insights on its interpretation}
\label{sect:discussion.size.luminosity}
In Section~\ref{sect:discussion.grain.growth} we have shown that, under the assumption of optically thin emission, the new 3\,mm observations of the brightest 35 Lupus discs 
can be readily explained in terms of fluxes, spectral indices, and reasonable dust masses with the presence of large millimetre-sized grains. 
An alternative explanation that does not require the presence of large grains has been extensively discussed \citep[e.g.,][]{Ricci:2012aa} and posits that the disc emission has a non-negligible optically thick contribution.

Lending support to this latter scenario is the correlation between the discs continuum size (the radius enclosing 68\% of the disc flux) and their luminosity $L_\mathrm{mm}\propto R_\mathrm{eff}^2$ \citep{Tripathi:2017aa,Andrews:2018ab} found in a sample of sub-arcsecond resolution observations at 0.89\,mm of 50 nearby protoplanetary discs in the Taurus-Auriga and Ophiuchus star-forming region.
The presence of such a relation has recently been confirmed also using ALMA observations \citep{Tazzari:2017aa,Andrews:2018ab,Hendler:2020aa,Sanchis:2020aa}. 
Simple dust evolution models \citep{Birnstiel:2012yg} are able to reproduce the observed trend for reasonable initial conditions
by assuming that grain sizes are set by radial drift \citep[see also][]{Rosotti:2019ab}.
However, \citet{Tripathi:2017aa} suggested that the observed relation can be alternatively explained if the discs are characterised by optically thick emission in narrow, spatially unresolved, regions, with filling factors (ratio of area covered by optically thick region to total disc area) of a few tens of percent. This would imply that the low spectral indices $\alphamm$ are determined by these optically thick regions rather than by the presence of genuinely large grains.  

Since \citet{Tripathi:2017aa} has only considered observations at 0.89\,mm, we aim to re-evaluate this latter conclusion in light of the new 3\,mm observations presented here. We focus our analysis on the IQ~Tau disc which is discussed in detail in \citealt{Tripathi:2017aa} and it is claimed that its emission can be explained with a 10-30\% optically thick emission within 80\,au. 
For the purpose of this simple modelling, let us assume that the over-density (hence, the optically thick emission) is concentrated at the \textit{core} of the disc rather than being distributed in small-scale structures throughout the disc extent: we write the dust surface density as $\Sigma_\mathrm{dust}(R)=\Sigma_0$ for $R\leq \Rcore$ and $\Sigma_\mathrm{dust}=0$ elsewhere. Although this may look less realistic than a myriad of localised rings emitting highly optically thick emission, the quantity of interest here is the total amount of optically thick contribution (and the mass it may be hidden in it) rather than its spatial distribution. 
The total disc mass is:
\begin{equation}
\begin{split}
    M_\mathrm{tot} &= 2\pi f \zeta \int_{R_\mathrm{in}}^{R_\mathrm{out}}\Sigma(R) R \mathrm{d} R = 2\pi f \zeta \int_{R_\mathrm{in}}^{\Rcore} \Sigma_0 R \mathrm{d} R = \\
      &= \pi \Rcore^2 \Sigma_0 f \zeta \simeq 3.515 \times 10^{-7} \left(\frac{\Rcore}{\mathrm{au}} \right)^2 \Sigma_0 f \zeta\   M_\odot\,,
\end{split}    
\end{equation}
where $f$ is the optically thick filling factor ($0\leq f\leq 1$), $\zeta$ is the gas-to-dust mass ratio, and we assumed that $R_\mathrm{in}\ll R_\mathrm{core}$.
If we assume the same opacity $\kappa_\mathrm{0.89\,mm}=3.5$ cm$^2$\,g$^{-1}$ used by \citet{Tripathi:2017aa}, a gas-to-dust mass ratio $\zeta=100$, and we take IQ~Tau's effective radius  as the core radius ($\Rcore=R_\mathrm{eff}=80\,$au), the disc mass would be $M_\mathrm{tot}\sim 6\times10^{-3}\,\MSun$, which we derived assuming a small filling factor $f=10\%$ and the minimum surface density (${\Sigma_0=0.285\,\mathrm{g\,cm}^{-2}}$) required to have an 0.89\,mm optically thick emission ($\tau_\nu=\kappa_\nu\Sigma_\mathrm{dust}=1$). 
Considering that IQ~Tau has an integrated flux of 170\,mJy, 
the median of the \citet{Tripathi:2017aa} sample, we conclude that half of the discs in that sample (which roughly includes the 50 brightest discs in the Taurus-Auriga and Ophiuchus regions) would have larger optically thick regions and, therefore, larger disc mass than IQ~Tau. For reference, for a disc with an integrated flux at 1\,mm of 1\,Jy, the size-luminosity relation predicts an effective size of 160\,au which would imply a total disc mass of $M_\mathrm{tot}\sim0.025\,\MSun$ for $f=10\%$. These disc masses are not unrealistically large at face value. However, they only represent the lower limit implied by the \citet{Tripathi:2017aa} argument: larger filling factors $f>10\%$ or larger optical depths $\tau_\nu>1$ would both require more massive discs.

Crucially, the additional information that we now have from our 3\,mm survey in terms of spectral indices suggest that the disc masses should be much larger than the nominal values that we have just derived, making the universal explanation of the size-luminosity relation in terms of optically thick substructures less straightforward. Indeed, the disc masses that we derived according to the arguments in \citet{Tripathi:2017aa} are the smallest compatible with their results, obtained assuming that the disc emission is marginally optically thick at 0.89\,mm ($\tau_\nu=1$). Assuming the disc is made of small grains ($\amax\ll 1\,$mm, $\beta\sim2$), if its core is marginally thick at 0.89\,mm, it would naturally become optically thin at longer wavelengths ($\tau_\nu\propto\nu^\beta$) and the disc would inescapably exhibit a large spectral index $\alphamm\gg 2$. In order to reproduce not only the disc integrated flux at 1\,mm, but also the typically observed $\alphamm\leq 2.5$, the core needs to be optically thick also at 1.3 and 3\,mm, thus implying an optical depth at 0.89\,mm, and thus disc masses, 5-10 times larger
($\tau_\mathrm{0.89\,mm}/\tau_\mathrm{3\,mm}=(3/0.89)^{\beta}\approx7$ for $\beta=1.7$)
than suggested by the \citet{Tripathi:2017aa} argument. We conclude that although the presence of optically thick regions can explain the size-luminosity relation on brightness grounds, it is much less viable in terms of required disc mass if the grains are small (i.e., with $\beta\sim2$). Indeed, while the $\alphamm$ values measured in the fainter fraction of discs can be well explained without the need of large grains, the low $\alphamm$ values measured in the bright and large discs require the genuine emission of large grains in order to yield realistic (i.e., not leading to gravitational instability) disc masses.

We emphasise that, although this argument does not disfavour the presence of optically thick substructures in general, it does show that they are not compatible with the observations (spatially-integrated mm fluxes and spectral indices) if the dust component is mostly made of small grains ($\beta\sim2$). The presence of widespread optically thick substructures made of large mm-sized grains (with large mm albedo, see \citealt{Zhu:2019aa}) still remains a viable scenario that will have to be investigated with multi-wavelength observations at higher resolution.

\section{Conclusions}
\label{sect:conclusions}
In this paper we have presented the first ALMA survey of protoplanetary discs at 3\,mm, targeting the 36 brightest Class~II protoplanetary discs in the Lupus star-forming region. 
The main results can be summarised as follows:
\begin{enumerate}[(1)]
    \item We obtained 3\,mm ALMA observations at $\sim0.35\arcsec$ resolution at a sensitivity between 20 and 50$\mu$Jy, which detected 35 out of 36 discs (Sz~91 was not detected due to an erroneous observational setup).
    \item By combining the new 3\,mm observations with previous ALMA observations at 0.89\,mm and a similar angular resolution, we find that all Lupus discs have spectral indices $\alphamm<3.0$, with a tendency of larger values for transition discs. The mean spectral index for the entire Lupus sample is $\alphamm\simeq 2.23\pm0.06$, while for the transition discs is $\alpha_\mathrm{TD}\simeq 2.5\pm0.1$. 
    \item Under the assumption of optically thin and Rayleigh-Jeans emission, the low spectral indices can be interpreted as evidence of large grains, with a median dust opacity power-law index $\beta\simeq0.23\pm0.06$, which require grains with $\amax>1\,$mm for a range of reasonable dust composition and porosity.
    \item We find that the distribution of spectral indices in Lupus is statistically indistinguishable from that of the Taurus and Ophiuchus star-forming regions, suggesting that dust retention mechanisms are common in discs from their early stages of evolution.
    \item The mean disc dust mass that we obtain from the 3\,mm fluxes is $22\pm2\,M_\oplus$, smaller by $\sim$20\% than those obtained from previous 0.89\,mm observations, but in line with those obtained from 1.3\,mm observations.
    \item We revisit the claim that the millimeter continuum size-luminosity relation can be explained by the widespread presence of localised substructures that emit optically thick radiation \citep{Tripathi:2017aa}. In light of the new 3\,mm measurements presented in this study we argue that the disc masses implied by such scenario would be possible, albeit very high if the grains were small. 
\end{enumerate}
In this paper we have focused on the spatially-integrated fluxes and spectral indices. A forthcoming paper will present the spatially-resolved study of the multi-wavelength Lupus disc observations \citep{Tazzari:2020ab}.

\section*{Acknowledgements}
We thank Cathie J. Clarke and Richard Booth for providing useful advice and Juan M. Alcal\'a for helpful comments on the manuscript. 
We are thankful to the anonymous referee for providing useful comments that improved the clarity of the paper.
This paper makes use of the following ALMA data: 
ADS/JAO.ALMA\#2016.1.00571.S.,  
ADS/JAO.ALMA\#2013.1.00220.S, 
ALMA is a partnership of ESO (representing its member states), NSF (USA) and NINS (Japan), 
together with NRC (Canada), MOST and ASIAA (Taiwan), and KASI (Republic of Korea), in 
cooperation with the Republic of Chile. The Joint ALMA Observatory is operated by 
ESO, AUI/NRAO and NAOJ.
M.T. has been supported by the UK Science and Technology research Council (STFC) via the consolidated grant ST/S000623/1, and by the European Union’s Horizon 2020 research and innovation programme under the Marie Sklodowska-Curie grant agreement No. 823823 (RISE DUSTBUSTERS project).
J.P.W. acknowledges support from NSF grant AST-1907486.
G.G. acknowledges the financial support of the Swiss National Science Foundation in the framework of the NCCR PlanetS.
N.M. acknowledges support from the Banting Postdoctoral Fellowships program, administered by the Government of Canada.
This research was partially supported by the Italian Ministry of Education, Universities and Research through the grant Progetti Premiali 2012 iALMA(CUP C52I13000140001), the Deutsche Forschungsgemeinschaft (DFG, German Research Foundation) - Ref no. FOR 2634/1ER685/11-1 and the DFG cluster of excellence ORIGINS (www.origins-cluster.de), from the EU Horizon 2020 research and innovation programme, Marie Sklodowska-Curie grant agreement 823823 (Dustbusters RISE project), and the European Research Council (ERC) via the ERC Synergy Grant {\em ECOGAL} (grant 855130).
This work was partly supported by the Deutsche Forschungs-Gemeinschaft (DFG, German Research Foundation) - Ref no. FOR 2634/1 TE 1024/1-1.

\section{Data availability}
\label{sect:data.availability}
The raw data used in this article is publicly available on the ALMA Archive (see project codes in the Acknowledgements). \tbref{tb:YSOs} and \tbref{tb:measurements.band3} are available in machine-readable format at \href{http://doi.org/10.5281/zenodo.4756282}{http://doi.org/10.5281/zenodo.4756282}.
The data underlying this article will be shared on reasonable request to the corresponding author.


\bibliographystyle{mnras}
\bibliography{mt_disks} 

\bsp    
\label{lastpage}
\end{document}